\documentclass[12pt]{article}
\begin{document}
\vspace*{2cm}
\thispagestyle{empty}
\begin{center}
{\Large  \bf Habitable sphere \\and fine structure constant,
$\alpha$}

\vspace{2cm} {\large Miroslaw Kozlowski,
Janina Marciak-Kozlowska\\
Institute of Electron Technology\\Al. Lotnik\'{o}w 32/46, 02-668
Warsaw, Poland}
\end{center}

\vspace{1.5cm}
{\def\baselinestretch{1}\hbox to 5 cm{\hsize=5cm\vbox{\ \hrule}}\par
$^*$Corresponding author, e-mail: MiroslawKozlowski@aster.pl
}
\vspace{2cm}

\begin{abstract} Future space missions, TPF and Darwin will focus
on searches of signatures of life on extrasolar planets. In this
paper we look for model independ definition of the habitable zone.
It will be shown that the radius of the habitable sphere depends
only on the constants of the Nature, $\alpha, \hbar, c$.

 {\bf Key
words:} Habitable sphere; Fine structure constant $\alpha$.

\end{abstract}
\newpage
\section*{Introduction}
The existence of Extra-Solar planets is well established. In the
contemporary status of the searching program (e.~g. DARWIN space
infrared interferometer project) the following categories of
extra-solar planets are described: Definite planets (20), possible
planets (8), microlensed planets (5), borderline planets (2), dust
clump planets (7) and pulsar planets (4), number in paranthesis
denotes the number of planet.\footnote{Data taken from
http://art.star.rl.ac.uk/darwin/planets} It is well known that
round the Sun the habitable zone -- Ecosphere exists. Within the
Sun Ecosphere are: Venus, Earth and Mars and Sun. It will be
interesting to
 calculate the
Ecosphere radius for ``average'' star with mass $M_s=a_G^{3/2}m_p$ ($a_G$ = the
gravity fine structure constant and $m_p$ = proton mass).

To that aim in this paper we investigate the possibility of the
calculations of the planet orbit radii as the function of the fine
structure constant $\alpha$. We argue that the Ecosphere is
defined as the part of space rounded the star which can be
calculated assuming the present day value of the electromagnetic
fine structure constant. Considering  the existence of the
,,niche'' for fine structure constant we calculate the niche for
planet orbit radii and obtain $R_{rel}=R(\alpha)/R(\alpha = 1/137)
= 0.5 - 1.5$ where $R_{rel}$ denotes the relative orbit radii. In
the case of the Solar system in Ecosphere we find out the orbits
of Venus, Earth and Mars. Considering the agreement of the
calculation with the Ecosphere radius for Solar system we argue
that our model for habitable zone can be applied to other planet
systems also.

\section*{Coincidence}
In paper~\cite{1} the quantum heat transport on the atomic, nuclear and quark
scale was discussed and the characteristic time scales were obtained. For
atomic scale:
\begin{equation}
\tau_a=\frac{\hbar}{m_e\alpha^2c^2},\label{eq1}
\end{equation}
where $m_e$ is electron mass, $\alpha$ is the electromagnetic fine structure
constant, $c$ is the light velocity. For nuclear scale:
\begin{equation}
\tau_n=\frac{\hbar}{m_n(\alpha^s)^2c^2},\label{eq2}
\end{equation}
where $m_n$ denotes nucleon mass, $\alpha^s\sim0.15$ is the coupling
constant for strong interactions. In the case of free quark gas (if it exist!):
\begin{equation}
\tau_q=\frac{\hbar}{m_q(\alpha^q_s)^2c^2}\label{eq3}
\end{equation}
and $\alpha^q_s, m_q$ are the fine structure constant for quark-quark
interaction and quark mass respectively.

The atomic time scale, $\tau_a$, is proportional to the ,,atomic year'', $T_a$,~viz.:
\begin{equation}
\tau_a=\frac{\hbar}{m_e\alpha^2c^2}\sim\frac{a_B}{\alpha \;
c}=T_a,\label{eq4}
\end{equation}
where $a_B$ is the Bohr radius.

It is quite interesting to observe that the ,,coincidence'' holds:
\begin{eqnarray}
A \; T_a&\sim& T_{\rm Earth},\label{eq5}\\
A \; m_p&=&1g,\nonumber
\end{eqnarray}
where $A$ is the Avogardo number, $A=6.02\,10^{23}$,
$m_p=1.66\,10^{-27}$ kg is the proton mass and $T_{\rm Earth}$
denotes the Earth year (in seconds).

From Kepler law the relation $T_{\rm Earth} \rightarrow R_{\rm
Earth}$ can be concluded:
\begin{equation}
T^2_{\rm Earth}=\left(\frac{2m_{\rm Earth} \pi}{(-m_{\rm
Earth}K)^{1/2}}\right)^2R^3_{\rm Earth}.\label{eq6}
\end{equation}
In formula~(\ref{eq6}) we approximate Earth orbit as the circle
with radius $R_{\rm Earth}$ and $m_{\rm Earth}$ is the Earth
mass,~$K$ is equal:
\begin{equation}
K=-Gm_{\rm Earth}M,\label{eq7}
\end{equation}
where $G$ is gravity constant and $M$ denotes the mass of the central body (the
Sun) which creates gravity forces. In the following we approximate the $M$ mass
of the central body by the mass of the ,,average'' star~\cite{2, 3}
\begin{equation}
M\cong a^{-3/2}_Gm_p=Nm_p,\label{eq8}
\end{equation}
where $N=a^{-3/2}_G$ is the Landau-Chandrasekhar number, $a_G$ denotes the fine
structure constant for gravity force. Comparing formulae~(\ref{eq5})
and~(\ref{eq6}) one obtains:
\begin{equation}
R^{3/2}=\frac{A\hbar c}{2\pi
\alpha^2}\frac{1}{m_ec^2}\left(\frac{M_{pl}}{m_p}\right)^{1/2}
\left(\frac{\hbar c}{m_p c^2}\right)^{1/2}.\label{eq9}
\end{equation}
In formula~(\ref{eq9}) for planet radius we omit the subscript ,,Earth''
because the radius does not depend on the planet mass. The $R$
denotes the planet orbit radius for average star with mass described by
formula~(\ref{eq8}). The planet radius depends only on the three fundamental
constant of the Nature: $G, \hbar, c$. The mass $M_{pl}=(\frac{\hbar
c}{G})^{1/2}$ is the Planck mass.

Considering formula~(\ref{eq5}), $A\; m_p$=1 g the planet radius~(\ref{eq9}) can
be formulated in more ,,elegant'' form:
\begin{equation}
R^{3/2}=\frac{\hbar c}{m_p
\alpha^2}\frac{1}{m_ec^2}\left(\frac{M_{pl}}{m_p}\right)^{1/2}\left(\frac{\hbar
c}{m_p c^2}\right)^{1/2}.\label{eq10}
\end{equation}
The dependence of $R$ on $\alpha$ is quite interesting. For, it is well known
that grand unified theories allow very sharp limits to be placed on the
possible vales of the fine structure constant in a cognizable universe. The
possibility of the doing physics on the background space-time at the
unification energy and the existence of stars made of protons and neutrons
endorse $\alpha$ in the niche~\cite{4}:
\begin{equation}
\frac{1}{180}\leq\alpha\leq\frac{1}{85}\label{eq11}
\end{equation}
or~\cite{5}
\begin{equation}
\frac{1}{195}\leq\alpha\leq\frac{1}{114}.\label{eq12}
\end{equation}
It is interesting to observe that one obtains the niche for planet
radii ---~the Ecosphere which is the consequence of
formulae~(\ref{eq11}) and (\ref{eq12}). The Ecosphere spans from
$R_{rel}\sim 0.5$ to $R_{rel}\sim 1.5$. In the case of the Solar
system in this niche we find only the orbits of Venus Earth and on
the border of the Ecosphere: Mars.

Considering the agreement of the calculations present in this
paper with the habitable zone for Sun, we argue that our model for
Ecosphere can be applied to other planet systems (other
``worlds'') also.

The habitable sphere is conservatively defined as a shell region
around a star where liquid water can be found at the surface of
the planet. this zone is typically from about $R_{\rm rel}$ 0.95
Au to 1.37 AU. As we shown this "life shell" can be calculated as
the function of the Nature constants $\alpha, \hbar, c$.

\end{document}